# Descriptor and Graph-based Molecular Representations in Prediction of Copolymer Properties Using Machine Learning

Elaheh Kazemi-Khasragh,[†,‡] Rocío Mercado,[*,¶] Carlos Gonzalez,[†,‡] and Maciej Haranczyk[*,‡]

†*Departamento de Ciencia de Materiales, Universidad Politécnica de Madrid, E.T.S. de Ingenieros de Caminos, 28040 Madrid, Spain*

‡ *IMDEA Materials Institute, C/Eric Kandel 2, 28906 Getafe, Madrid, Spain*

¶*Department of Computer Science and Engineering, Chalmers University of Technology, Chalmersplatsen 4, 412 96 Gothenburg, Sweden*

E-mail: rocio.mercado@chalmers.se; maciej.haranczyk@imdea.org

## Abstract

Copolymers are highly versatile materials with a vast range of possible chemical compositions. By using computational methods for property prediction, the design of copolymers can be accelerated, allowing for the prioritization of candidates with favorable properties. In this study, we utilized two distinct representations of molecular ensembles to predict the seven different physical polymer properties copolymers using machine learning: we used a random forest (RF) model to predict polymer properties from molecular descriptors, and a graph neural network (GNN) to predict the same properties from 2D polymer graphs under both a single- and multi-task setting. To train and evaluate the models, we constructed a data set from molecular dynamic simulations for 140 binary copolymers with varying monomer compositions and configurations. Our results demonstrate that descriptors-based RFs excel at predicting density and specific heat capacities at constant pressure ($C_p$) and volume ($C_v$) because these properties are strongly tied to specific molecular features captured by molecular descriptors. In contrast, graph representations better predict expansion coefficients ($γ$, $α$) and bulk modulus (K), which depend more on complex structural interactions better captured by graph-based models. This study underscores the importance of choosing appropriate representations for predicting molecular properties. Our findings demonstrate how machine learning models can expedite copolymer discovery with learnable structure-property relationships, streamlining polymer design and advancing the development of high-performance materials for diverse applications.

## Introduction

Polymers play an integral role across diverse industries, driven by their abundant accessibility and adaptable properties. From structural materials in construction to biodegradable packaging solutions and advanced medical applications, polymers exhibit remarkable versatility, catering to a wide spectrum of applications in industries worldwide.[1–4] To leverage the vast usage of polymers in industry, there is a crucial need to precisely control the properties of these materials according to specific applications, whether it be optimizing strength for mechanical requirements or tailoring thermal resistance based on industry preferences. The challenge lies in the complexity of characterizing and optimizing diverse polymer properties to meet specific demands. The complexities in characterizing these materials, often due to their diverse properties, hinder traditional experimental design approaches which tend to be resource-intensive and time-consuming.[5–7]

There is a need for improved methods for the design of polymer materials which meet unique industry requirements. These requirements may include enhanced mechanical properties or thermodynamic properties to increase stability,[8,9] optimizing energy efficiency,[10] ensuring material integrity,[8] or minimizing safety risks, [11] depending on the material's end- use. These requirements are essential for achieving effective system performance for a target application.

In addition to the complexities associated with characterizing new systems, another bottleneck arises from the large chemical and configuration space accessible to polymers. Consider a linear copolymer chain composed of $n$ potential monomers and $m$ chemical moieties. The total number of unique sequences that can be expressed in this copolymer can be expressed as $m^n/2$.[12] The division by 2 in the exponent accounts for eliminating the double-counting of polymer sequences, where a sequence and its reverse represent the same copolymer. For instance, in the case of an AB-type copolymer with two chemical moieties and a chain length of 50, the total combinations amount to $2^{49}$, surpassing $10^{15}$. Consider that many copolymers possess more than two chemical moieties and several hundreds of monomer units, so navigating through this expansive space is like searching for a needle in a haystack.[12–15] To address this complexity, efficient tools are essential for navigating through this chemical space and to minimize the number of property measurements needed to complete a design task within a reasonable time-frame and budget.

Polymer design has evolved from traditional empirical methods, relying on trial-and- error experimentation, to more advanced strategies. [16] Computational methods, including computer-aided design and simulation, have enabled a systematic exploration of the polymer chemical space, allowing for more precise predictions of properties. Combination chemistry and high-throughput screening have accelerated polymer discovery by enabling the simultaneous synthesis and testing of large polymer libraries. In recent years, materials informatics and machine learning (ML) have further transformed polymer design, leveraging data- driven models to predict properties and streamline the identification of promising materials.

Overall, these evolving strategies have enhanced the efficiency and precision of polymer design,

moving from intuition-driven approaches to more systematic, technologically advanced methodologies.[17–21]

Numerous works have explored various strategies for predicting polymer properties, using data-driven approaches. The successful utilization of neural networks [22,23] and support vector regression[24] in predicting glass transition temperature, tribological properties, electronic and dielectric properties, storage modulus, damping, specific heat capacity, and mechanical characteristics, underscores the immense potential of ML methods in this domain.[25,26] In our two recent studies, ML played a crucial role in predicting the physical properties of homopolymers.[27,28]

More recently, ML-based methods have made significant progress in screening polymer libraries. For instance, AlFaraj et al. [29] and Gurnani et al. [30] demonstrated the effectiveness of ML in accelerating feature extraction and enabling the large-scale screening of massive polymer libraries.

To enable ML-driven predictions, researchers are particularly interested in creating ex- tensive data sets, often in conjunction with simulation methods. For instance, Jang et al. [31] developed a comprehensive database of 789 epoxy resins, detailing four crucial properties (density, coefficient of thermal expansion, glass transition temperature, and Young's modulus) derived from molecular dynamics (MD) simulations. Tao et al. [32] harnessed MD simulations to construct a vast data set on polymer fractional free volume for >6500 homopolymers and 1400 polyamides. Using ML models, they established composition-structure relation- ships, surpassing traditional group contribution theories with efficient feed-forward neural network (FFNN) models.

A substantial number of different molecular representations have been developed to enhance system descriptions of polymers in the field of ML. One of the most important molecular representations are molecular graphs, and several studies have highlighted the use of graph neural networks (GNNs) in the field of polymer property prediction. [33,34] Zeng et al. [35] emphasized the significance of graph convolutional neural networks (GCNN) in predicting polymer properties, showing remarkable agreement with density functional theory (DFT) results and outperforming other ML algorithms. Graph-based approaches remove the need for complex hand-crafted descriptors while maintaining prediction accuracy. Recently, Aldeghi and Coley [36] introduced a graph representation of molecular ensembles and a GNN architecture tailored for polymer property prediction. Their work demonstrated that using this framework to model polymers, they were able to capture the relevant features that distinguish polymer materials from one another, outperforming traditional cheminformatics methodologies for polymer representation.

These studies in polymer property prediction showcase the transformative impact of ML, which has empowered researchers to leverage diverse techniques and expansive data sets for more accurate polymer property prediction. The integration of ML with simulation methods and advanced molecular representations signals a promising direction for advancing our understanding and predictive capabilities in polymer science.

Here, we present a data-driven pipeline for predicting copolymer properties. To address the lack of available experimental measurements for copolymer systems, we used MD simulations to compute properties for 140 copolymers and prepared an ML-ready data set for building predictive models from MD data. Given the complexity of copolymer structures, representing them compactly for ML models is a challenge. To tackle this, we explored two representations: descriptor fingerprints constructed from the monomers of each copolymer and a polymer graph representation. [36] These representations serve as inputs to random forest (RF) and graph neural network (GNN) models, respectively, and are used to predict the computed MD properties in both single- and multi-task settings, enabling the prediction of copolymer properties at an MD-level of accuracy but at a fraction of the computational cost. In this study, we investigated three distinct classes of copolymers: alternating, random, and block. Our findings demonstrate that MD can accurately model the properties of all copolymer classes in our data set when compared to the limited experimental data available.

We also demonstrate that ML models trained on MD data in both single- and multi-task settings can accelerate the prediction of copolymer properties, with multi-task ML models generally outperforming the single-task methods for most properties.

## Methods

### Experimental data curation

We extracted 146 reference values from the PolyInfo[37] database for comparison to the copolymers studied in this work. The extracted values corresponded to the 126 copolymers studied in this work which had experimentally measured values for density ($\rho$; 68 points), $C_p$ (specific heat capacity at constant pressure; 48 points), $\gamma$ (volume expansion coefficient; 11 points), $\alpha$ (linear expansion coefficient; 5 points), and K (bulk modulus; 14 points). Due to access restrictions policies set by the database administrators, only a limited number of data points were able to be downloaded from the PolyInfo database.

### Molecular dynamics

We used MD simulations to model various copolymer properties at the molecular level. Using the Large-scale Atomic/Molecular Massively Parallel Simulator (LAMMPS)[38] through the RadonPy interface,[39] we ran MD simulations of copolymer chains using the General Amber Force Field (GAFF).[40]

We start with the generation of the initial unit cell structures. A polymer chain is constructed by connecting a repeating unit consisting of 2 monomers via the self-avoiding random walk algorithm.[39] We equilibrate the system through a meticulous 21-step compression/decompression equilibration protocol proposed by Larsen and co-workers.[41] This protocol orchestrates temperature cycling from 600 K to 300 K, coupled with compression (50,000 atm) and decompression (1 atm) steps. Temperature and pressure are regulated through NVT and NPT simulations using a Nosé–Hoover thermostat and barostat.

After the 21-step equilibration protocol, NPT simulations are run for each system at 300 K and 1 atm until equilibrium is achieved (typically more than 5 ns). Following equilibrium, an extensive suite of copolymer properties is calculated, including $\rho$, $R_g$, $C_p$, $C_v$, $K$, $\alpha$, and $\gamma$. The $R_g$ is a measure of the spatial extent of the polymer chain, defined as the root mean square distance of the polymer's atoms from its center of mass. Full details are provided in the Supporting Information.

The calculated properties were compared with experimental data extracted from the PolyInfo database[37] to validate the accuracy of the calculations. We analyzed the computed properties via the following statistical metrics: squared correlation coefficient ($R^2$), Spear- man's rank correlation ($\rho_s$), and mean squared error (MSE). These metrics collectively assess the linear agreement between predicted and experimental values, as well as the magnitude of the error in the predictions, providing a comprehensive evaluation of model accuracy. Additionally, we generated parity plots to visually inspect model performance and potential biases.

## ML-ready data set

Using MD, we generated a data set of 140 copolymers and their properties. For each copolymer, we computed its $\rho$, $R_g$, $C_p$, $C_v$, $K$, $\alpha$, and $\gamma$. This resulted in a final data set size of 980 entries, comprising 92 random copolymers, 8 block copolymers, and 40 alternating copolymers, each using different types of monomers. In the general structure of the copolymers, there are 97 different and unique monomers. Specifically, monomer A can be one of 45 unique monomers, while monomer B can be one of 57 unique monomers. There are 5 over- lapping monomers between the sets of monomers A and B. For full details on the monomers explored, please see the Supporting Information.

For training, validation, and testing of single-task models, the copolymer data set was split using 80%:10%:10% random splits. This random splitting ensures that each subset contains a representative sample of the overall dataset.

## PaDEL descriptor fingerprints

We represent each copolymer building block using PaDEL descriptors. These are computed from the SMILES representation of each copolymer building block using the PaDEL-Descriptor software.[42] PaDEL-Descriptor generates >1400 molecular descriptors, including 1D, 2D, and 3D descriptors. These descriptors have been previously demonstrated to accurately model the properties of interest in similar systems.[6,27,43] To compute the descriptor fingerprint, $D$, for a copolymer, we sum the monomer descriptor fingerprints in proportion to their composition in the copolymer. It is crucial to highlight that $D$ is a property that depends solely on the fractional composition of the individual monomers, irrespective of their distribution or arrangement within the copolymer. This means that the calculated descriptor fingerprint $D$ reflects the proportion of each monomer type rather than their specific spatial distribution within the polymer chain. For example, for copolymers comprising two monomer components, this can be expressed as

$$D = c_1 \cdot D_1 + c_2 \cdot D_2,$$

where $c_1$ and $c_2$ represent the fraction of each monomer in the copolymer, and $D_1$ and $D_2$ are sets of descriptors characterizing the molecular properties of the individual monomers. This approach enables us create a vector representation for a copolymer from the descriptor fingerprints of its individual building blocks. The input descriptors were standardized using min-max normalization before training. This step ensures that all features contribute equally to the model training process by scaling each feature to a given range, typically [0, 1]. It means that, for every descriptor, the minimum value of that descriptor gets transformed into a 0, the maximum value gets transformed into a 1, and every other value gets transformed into a decimal between 0 and 1.[44]

## Copolymer graph representation

One limitation of the descriptor fingerprints is that they cannot encode for differences in how the monomers are connected, e.g., whether in a random, block, or alternating copolymer structure (Figure 1). To tackle this limitation, we use a graph representation of copolymer structure that allows us to encode more complex information about monomer connectivity.[36,45]

In this graph representation, atoms are represented as vertices and bonds as edges, with each edge assigned a weight between 0 and 1 that reflects the probability of the bond appearing in each repeating unit. Through the connection of disparate monomers by edges, we can capture not only the recurring patterns inherent in polymer chains but also the spectrum of potential structural arrangements.

In AB binary alternating copolymers, the chain sequence follows an A-B pattern. The two ends of the repeating unit are connected, so in our representation all edges have a weight of 1. In random AB copolymers, a variety of arrangements can emerge, encompassing A-A, A-B, and B-B connections. In contrast, block copolymers maintain these patterns, yet A-A and B-B connections prevail over A-B connections in frequency. This discrepancy is reflected in the assigned weights. In the random configuration, uniform weights of 0.5 are assigned to A-A, B-B, and A-B connections. Conversely, in the block configuration, A-A and B-B connections are notably higher at 0.95, while A-B connections are markedly lower at 0.05.

## Random forest model

We use a random forest (RF)[46] as a baseline to predict the properties of interest from the PaDEL descriptor fingerprints. The RF allows us to efficiently uncover the relation- ships between the copolymer's structural attributes and its properties, thus advancing our understanding of copolymer behavior and performance. RFs have been previously shown to be effective in numerous predictive tasks involving molecules and materials.[27,47,48] For the RF implementation, we use the scikit-learn[49] library.

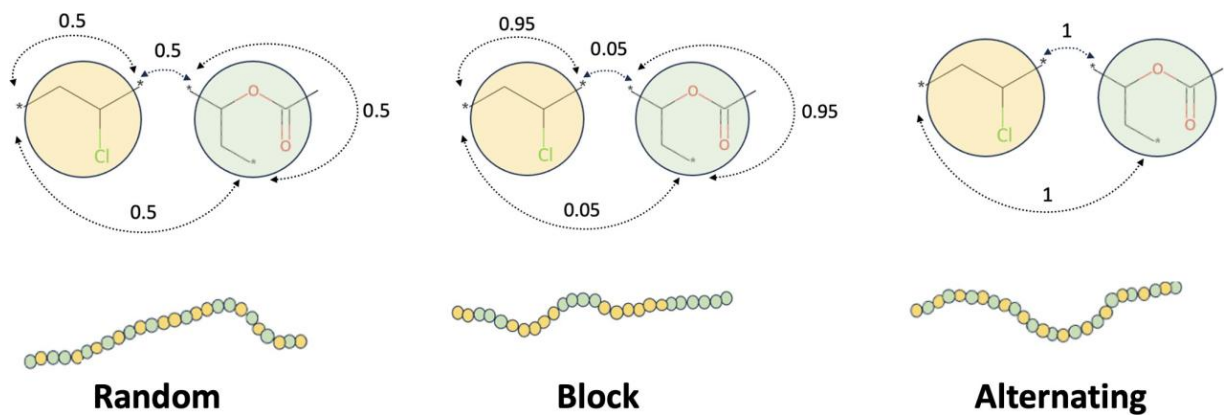

Figure 1: Schematic of the graph representation used in this work for random, block, and alternating copolymers.

We train the models using the mean-squared error (MSE) loss function and the parameters listed in Table 1. For all other parameters, we used the default values. We use the MSE loss function because it effectively penalizes larger errors, providing a smooth gradient that facilitates stable and efficient model training. MSE helps the model adjust more precisely to significant deviations in the data, enhancing overall accuracy.[50]

To address the multi-task learning scenario, where multiple properties are predicted simultaneously, we adapted the RF model to a multi-task setting. Multi-task learning is particularly beneficial when the tasks (in this case, property predictions) share common underlying structures, as it allows the model to leverage shared information across tasks, potentially improving performance on each individual task.

While multi-task RF is not directly available as a standard library function, we implemented this by modifying the scikit-learn RandomForestRegressor to handle multiple output targets simultaneously. In scikit-learn, this can be achieved by providing a matrix of tar- get variables, where each column corresponds to a different task (i.e., property to predict). The model is then trained to minimize the combined MSE across all tasks. For the multi- task RF, we specifically used the following hyperparameters: *n_estimators* was set to 100, *max_depth* was set to 10, *min_samples_split* was set to 4, *min_samples_leaf* was set to 2, and *max_features* was set to 'sqrt'.

By using this modified RF approach, we aim to capture the multi-task nature of the problem effectively, ensuring that the model benefits from the shared information between tasks. All model training and evaluation were conducted using scikit-learn, and the multi-task capability was handled natively within the library by structuring the target data appropriately.

Table 1: Hyperparameter values for the single-task RF models.

| Model | n_estimators | max_depth | min_samples_split | min_samples_leaf | max_features |
|---|---|---|---|---|---|
| $\rho$ | 100 | 20 | 2 | 1 | 'sqrt' |
| $R_g$ | 200 | 10 | 2 | 1 | 'log2' |
| $C_p$ | 500 | 80 | 2 | 1 | 'sqrt' |
| $C_v$ | 500 | 20 | 2 | 1 | 'sqrt' |
| K | 250 | 20 | 12 | 1 | 'sqrt' |
| $\alpha$ | 500 | 80 | 2 | 1 | 'sqrt' |
| $\gamma$ | 200 | 20 | 2 | 1 | 'sqrt' |

To evaluate the importance of each PaDEL descriptor, we used the RF model trained on the training set to compute feature importance. This was done using the mean decrease in impurity, as implemented in the scikit-learn[49] package.

## Neural network model

The network architecture utilized in this study is an extension of the directed message passing neural network (D-MPNN) implemented by Aldeghi and Coley [36], known as the weighted D-MPNN, or wD-MPNN. By weighting the edges based on the specific copolymer configuration and incorporating stoichiometry information, the wD-MPNN learns a more nuanced and effective representation of polymer structures compared to the D-MPNN.

The D-MPNN is a class of GNN that operates on molecular graphs. The process begins with the assignment of feature vectors to each node and edge, describing the properties of the corresponding atoms and bonds. These feature vectors encode information such as atom type and formal charge. Through iterative message passing, information is exchanged between neighboring atoms via directed edges, with each edge transmitting a message to update the feature vectors of adjacent nodes.

This iterative process allows information to propagate through the graph, refining the feature vectors based on directional relationships between atoms. After multiple iterations, the feature vectors are aggregated to produce a comprehensive representation of the molecule, which is subsequently fed into a FFNN for predicting molecular properties.[36,51,51–55]

The model was trained using the MSE loss as implemented in scikit-learn. The aggregation function for message passing in the models was set to 'sum', and the number of warm-up epochs were set to 10 . Three folds were used for cross-validation. Hyperparameters were optimized as outlined in the section below, with the parameters used for wD-MPNN models outlined in Table 2. For all other parameters, we used the default values in the Aldeghi and Coley [36] implementation.

Table 2: Hyperparameter values for the wD-MPNN models.

| **Model** | *epochs* | *depth* | *dropout* | *learning rate* | *ffn_num_layers* | *FFNN width* |
|---|---|---|---|---|---|---|
| $\rho$ | 500 | 6 | 0.5 | 1e-6 | 3 | 2200 |
| $R_g$ | 250 | 8 | 0.4 | 1e-6 | 4 | 2100 |
| $C_p$ | 500 | 6 | 0.5 | 1e-5 | 3 | 1800 |
| $C_v$ | 500 | 2 | 0 | 1e-6 | 3 | 700 |
| K | 250 | 5 | 0 | 1e-5 | 2 | 1700 |
| $\alpha$ | 250 | 7 | 0.1 | 1e-6 | 5 | 700 |
| $\gamma$ | 250 | 3 | 0.3 | 1e-5 | 3 | 1900 |

## Hyperparameter optimization

Hyperparameter optimization was performed for the RF models using a grid search.[56] The parameters tuned included: the number of estimators, sampled from [30, 50, 100, 200, 500]; the maximum depth of the trees, sampled from [10, 20, 80]; the maximum number of features considered for splitting a node, either 'log2' or 'sqrt'; the minimum number of samples required to split an internal node, adjusted between 2 and 12; and the minimum number of samples required to be at a leaf node, adjusted between 1 and 5. The wD-MPNN model parameters were optimized using the Optuna framework.[57]

Each wD-MPNN was tuned for 100 iterations using various parameters to optimize its performance according to the MSE objective function. The tuned parameters comprised of: the number of epochs, sampled from [100, 250, 500]; the MPNN *depth* (number of message passing steps), adjusted between 3 and 10 layers; the dropout rate, sampled between 0 and 0.5; the learning rate, adjusted between 1e-3 and 1e-6; and the number of layers in the final FFNN, set between 1 and 10. The FFNN width was set between 500 to 2500, with the ReLU activation function used and no regularization applied. The optimizer used was Adam. Additionally, the objective in Optuna was to minimize the MSE of the model predictions. By doing so, we aimed to find the set of hyperparameters that would result in the most accurate predictions for our polymer property datasets.

### Evaluation metrics

We used the Spearman's rank correlation ($\rho_s$), mean squared error (MSE) and squared coloration coefficient ($R^2$) to evaluate the performance of the RF and wD-MPNN models. The predictions made by each model were compared to the computed values from MD. The standard error was computed for all predictions using an ensemble model approach, where three separate models were trained using a different initialization, and the error computed based on these predictions.

## Results and discussion

### Molecular dynamics predictions versus experiment

Figure 2 illustrates a comprehensive comparison between simulated values and experimental values from the PolyInfo[37] data base of key thermodynamic properties for the copolymers under investigation. Additionally, Table 3 presents the $R^2$, Spearman's rank correlation coefficient ($\rho_s$), and MSE values corresponding to the simulated properties, offering further insights into the quality of the simulations.

For the density $\rho$, thermal-expansion coefficient $\alpha$, and surface tension $\gamma$, $R^2 > 0.70$, confirming that the simulations capture the overall experimental trend. The bulk modulus $K$ and heat capacity $C_p$ follow the same trend, albeit with lower $R^2$ values of 0.61 and 0.53, respectively. Inspection of Figure 2 further shows that

MD reproduces $C_p$ accurately for values below $\sim$3000 J kg$^{-1}$ K$^{-1}$; beyond this threshold the scatter widens and the simulations systematically underestimate the measurements.

The resulting fan-shaped residual pattern is a textbook manifestation of *heteroscedasticity*: the error variance grows with the magnitude of $C_p$. In this context the effect most plausibly originates on the experimental side; calorimetric $C_p$ measurements for compositionally complex copolymers are sensitive to sequence distribution, micro-phase segregation, and trace solvent content, factors that become increasingly influential at higher $C_p$. Limited data availability in PolyInfo amplifies this scatter. Hence, the heteroscedasticity should be viewed less as an intrinsic flaw of the MD force field than as an uncertainty envelope imposed by the reference data.

Despite these systematic deviations, the rank ordering remains reliable: every $\rho_s$ exceeds 0.60 and, for four of the five properties, surpasses 0.80. The present MD protocol thus provides a sound basis for screening studies that hinge on relative comparisons, while simultaneously underscoring the need for more precise experimental benchmarks when quantitative predictive accuracy is required.

Table 3: Simulated properties and their corresponding $R^2$, Spearman's rank correlation coefficient ($\rho_s$), and MSE values compared to experiment.

| Simulated Properties | $R^2$ | $\rho_s$ | MSE |
|---|---|---|---|
| Density ($\rho$) | 0.858 | 0.89 | 0.01 |
| $C_p$ | 0.53 | 0.83 | 0.53 |
| Bulk modulus (K) | 0.691 | 0.83 | 0.22 |
| Linear expansion coefficient ($\alpha$) | 0.965 | 0.99 | $6.4 \times 10^{-9}$ |
| Volume expansion coefficient ($\gamma$) | 0.705 | 0.60 | $3.4 \times 10^{-8}$ |

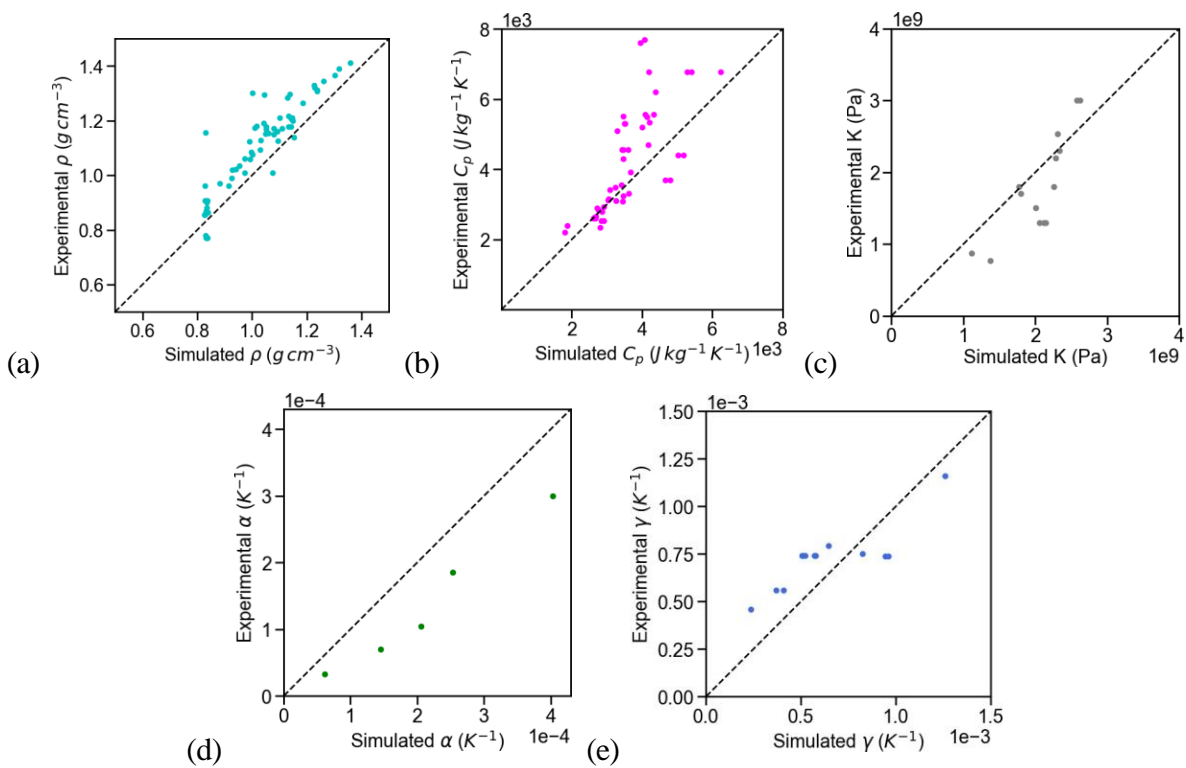

Figure 2: Experimental and simulated values of the (a) density, (b) $C_p$, (c) bulk modulus, (d) linear expansion coefficient, and (e) volume expansion coefficient.

## Machine learning

### Random forest model evaluation

In this section, we present results from testing the ability of single and multi-task RF models, using molecular descriptors obtained from the PaDEL-Descriptor toolkit, to predict the properties of copolymers. Figure 3a presents the $R^2$ values for the training, validation, and test sets across various properties, including $\rho$, $C_p$, $C_v$, $K$, $\alpha$, and $\gamma$.

The figure demonstrates the predictive performance of the single-task RF models developed using all calculated descriptors. Among the properties, density, $C_p$, and $C_v$ exhibit the highest $R^2$ values across all data sets, indicating strong predictive capabilities and suggesting that the RF models can accurately capture the underlying patterns for these properties. In contrast, the linear expansion coefficient and volume expansion coefficient show comparatively lower $R^2$ values, particularly in the validation and test sets, implying that the models struggle to generalize well for these properties.

The bulk modulus and $R_g$ present intermediate performance, with reasonable $R^2$ values but still lower than those of density, $C_p$ and $C_v$. These results highlight the varying efficacy of the RF models depending on the specific property being predicted.

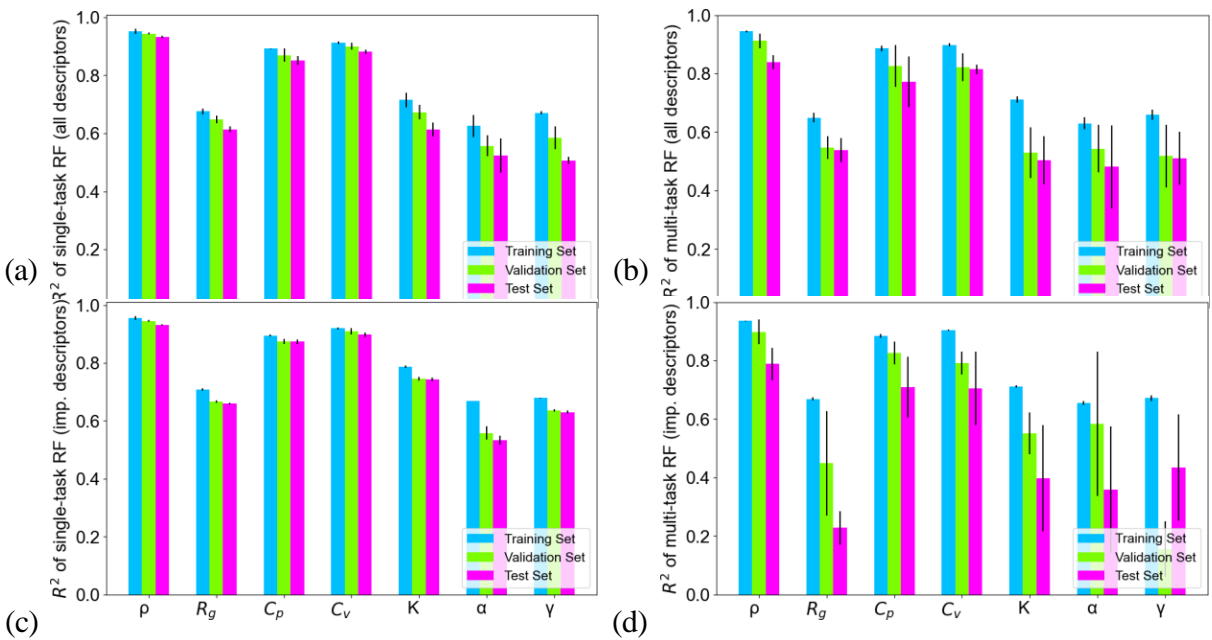

Figure 3: $R^2$ values for (a) single-task RF models using all descriptors, (b) multi-task RF models using all descriptors, (c) single-task RF models using the 10 most important descriptors, and (d) multi-task RF models using the 10 most important descriptors.
Results are presented for the training, validation, and test splits across all properties.

In contrast to the single-task RF models, the multi-task RF models show reduced $R^2$ values for all properties. Figure 3b presents the $R^2$ values for the training, validation, and test sets across all properties. The figure illustrates that the multi-task RF model generally achieves lower $R^2$ values compared to the single-task models across all properties. This reduction suggests that while multi-task modeling may offer efficiencies in computation and feature utilization, it often sacrifices predictive accuracy compared to single-task models tailored to specific properties.

We utilized feature importance analysis to reduce the number of descriptors. In Figure 4,

we present the top 3 most important descriptors for each model, selected using the training set. These analyses help streamline our RF models by focusing on the most influential descriptors, enhancing predictive performance and model interpretability. We retrained the RF models using only the top 10 most important descriptors coming from this analysis, and the updated results are presented in Figure 3c. Notably, in Figure 3, we observe a drop in performance on the test set for several properties. This suggests that while the selected descriptors are highly predictive for the training data, they may not generalize as well to unseen data. This highlights a potential limitation of selecting features solely based on training set importance, and suggests that incorporating validation or cross-validation-based feature selection might improve generalization.

Several molecular descriptors appear repeatedly across the top descriptors for various physical properties of copolymers, highlighting their significance in capturing diverse molecular interactions and structural attributes.

Through this analysis, we were able to identify the key molecular features influencing different properties in our random forest model, unveiling insightful structure-property relation- ships. For instance, we found that, unsurprisingly, resonance and charge distribution-related descriptors (e.g., TDB3r, AATS0v, AATS0m) are crucial for predicting thermal properties such as heat capacities ($C_p$ and $C_v$). In contrast, electronic and topological features (e.g., ETA_EtaP_F_L, TDB2s, SIC2) affect density and bulk modulus, possibly by providing information on how copolymer chains pack and respond to pressure. Additionally, descriptors such as BCUTw_1h, SpDiam_D, ATSC1e, and AATS6v capture nuanced molecular attributes—including hydrophobicity, molecular size, polarizability, and the spatial distribution of atomic properties— and as such these features were found essential for accurate property prediction. Finally, polarity and hydrogen-bonding features (e.g., MLFER_BO) play a pivotal role in predicting expansion behavior and the radius of gyration ($R_g$), as well as capturing key interactions relevant to thermal properties.

We also identified several molecular features that did not significantly contribute to the prediction accuracy of our random forest model. For instance, simpler features related to basic molecular counts and connectivity were found to be less influential. This lack of importance could be attributed to their inability to capture the complex interactions and nuanced structural properties that are critical for predicting the targeted physical properties. Additionally, basic geometric parameters did not provide substantial predictive power, possibly because they do not account for the intricate electronic and topological features that in- fluence properties like density and bulk modulus. These findings highlight the necessity of using more sophisticated and context-specific features to accurately model complex material properties.

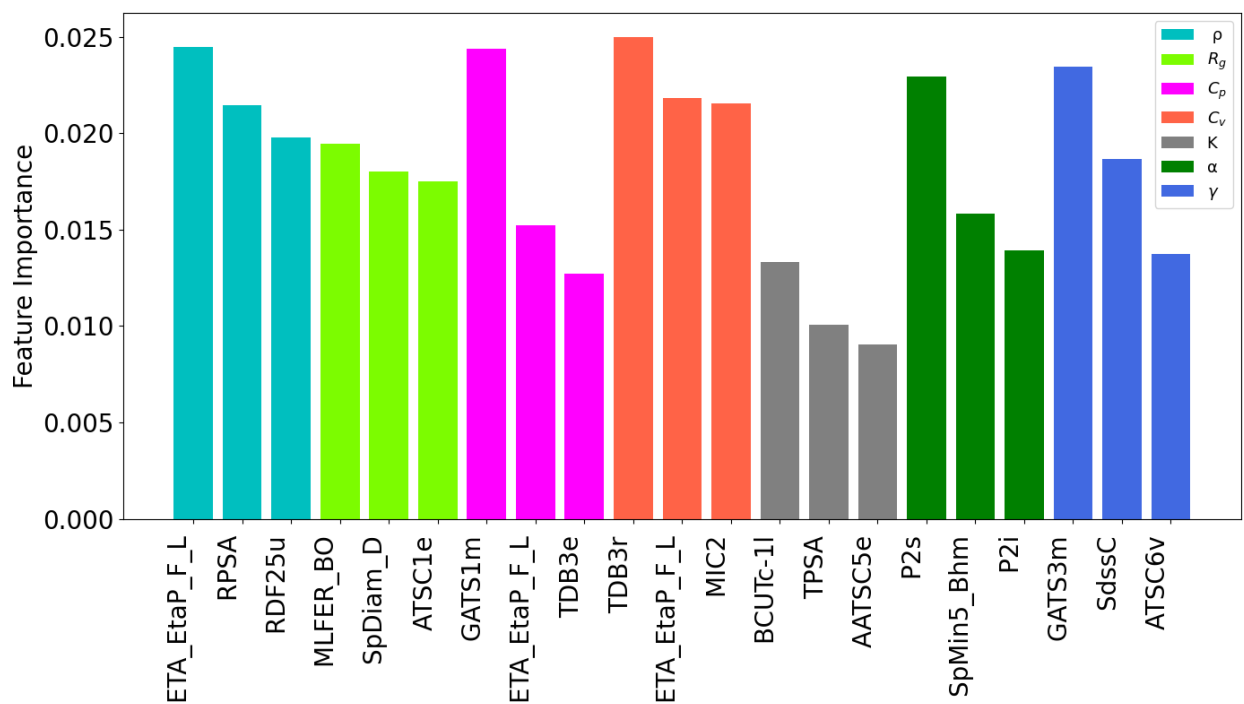

Figure 4: Important feature analysis showing the top 3 descriptors for each property ($\rho$, $R_g$, $C_p$, $C_v$, K, $\alpha$, and $\gamma$).

Given these insights, we constructed new single-task RF models with the 10 most important descriptors for each individual model to further refine our predictions. Figure 3c displays the $R^2$ values for these models. Remarkably, all the new single-task RF models demonstrated an increase in $R^2$ values compared to previously-discussed models, illustrating how a reduced feature space can enable a model to better learn the chemically meaningful variability in the data while minimizing learning from noise. Figure 3d illustrates the $R^2$ values of multi-task RF models predicting the same properties with the 10 most important descriptors. Interestingly, these models exhibit worse performance than multi-task models utilizing all descriptors. This is because in single-task models, removing less relevant descriptors reduces complexity and helps a model avoid overfitting to spurious correlations; in contrast, a fuller feature set can benefit a multi-task model as it can learn from auxiliary features that capture correlations between tasks. Since multi-task models rely on shared features across different tasks, reducing the descriptors to only the top 10 per property likely eliminated crucial features that would have helped the models better generalize between different tasks. Additionally, multi-task models benefit from the diversity of descriptors as they can leverage shared information across different properties, enhancing their ability to capture complex interdependencies and improve overall prediction accuracy. By limiting the feature set, we inadvertently constrained the model's capacity to utilize these beneficial cross-task correlations, leading to a decline in performance.

To summarize, this study explored single-task RF models using both the full descriptor set and a refined subset of important descriptors, revealing that feature reduction generally improved performance. In contrast, multi-task RF models performed better with the full descriptor set than with a reduced subset, highlighting the importance of feature diversity in multi-task learning. Overall, single-task RF models demonstrated superior predictive accuracy across all studied properties.

**Neural network model evaluation**

In this section, we showcase the results from evaluating the performance of wD-MPNNs models that utilize molecular graph representations to predict the properties of copolymers. Figure 5 presents the $R^2$ values for the training, validation, and test sets of the wD-MPNN models, predicting the same properties as those predicted with the RF models. The results indicate that for some properties, the wD-MPNN models achieve good accuracy, highlighting the effectiveness of the graph representation. However, the error in this representation stems from the inherent randomness in the design of random copolymers, as the randomness can vary significantly across different molecular dynamics (MD) designs.

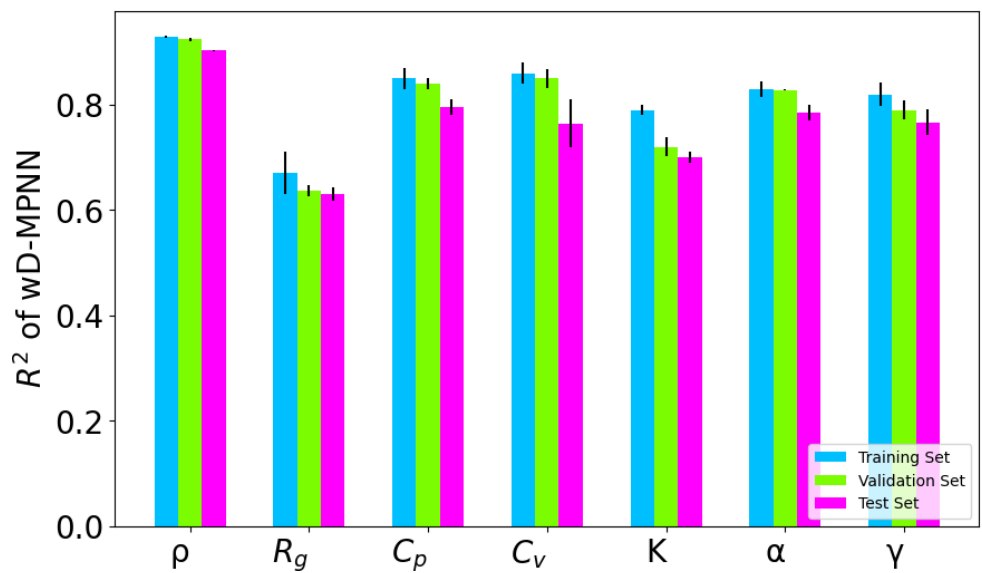

Figure 5: $R^2$ values of wD-MPNN models for the training, validation, and test sets across different properties.

We compared the predictive power of RF models based on descriptors with wD-MPNN models based on graph representations. Figure 6 compares the $R^2$ values of the test sets for various models: single-task RF with all descriptors, single-task RF with important descriptors, multi-task RF with all descriptors, multi-task RF with important descriptors, and wD- MPNN. This comparison highlights the predictive power of RF models based on descriptors and wD-MPNN models based on graph representations for predicting different copolymer properties. For instance, by using directed edges, D-MPNN models such as wD-MPNN can capture nuanced relational information about atoms, leading to improved representations and thus accuracy in molecular property prediction tasks.[36,51,51–55]

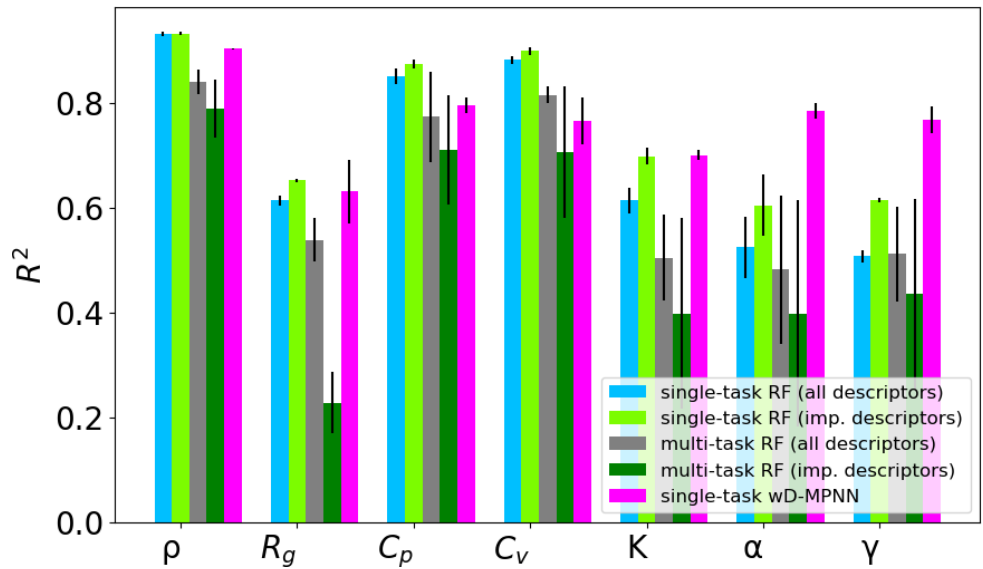

Figure 6: Comparison of $R^2$ values for the test sets across different models.

**Evaluating random forests versus neural networks**

Here, we compare the performance of RF models and wD-MPNN models in predicting 7 different properties of copolymers. By leveraging different ML techniques, we aim to determine the most effective method for capturing the intricate relationships between the structural at- tributes of copolymers and their resulting properties. This evaluation provides insights into the strengths and weaknesses of each approach, guiding the selection of the most suitable model for future predictive tasks in polymer science.

The RF models demonstrate better predictive performance for properties such as density, $C_p$, and $C_v$. For instance, the $R^2$ for density is 0.93 with single-task RF using only important descriptors, compared to 0.90 for the wD-MPNN, showing an improvement of approximately 3.3%. Similarly, for $C_p$ and $C_v$, the $R^2$ values are 0.87 and 0.89 for RF models, respectively, compared to 0.795 and 0.76 for the wD-MPNN, resulting in improvements of about 9.4% and 17.1%, respectively. This can be attributed to the effectiveness of molecular descriptors in capturing the specific characteristics relevant to these properties, as well as the relatively straightforward relationships between these properties and molecular structure.

On the other hand, the wD-MPNN models demonstrate superior predictive performance for properties such as $\alpha$, $\gamma$, and K compared to the RF models. For example, the $R^2$ for $\gamma$ is 0.76 for the wD-MPNN, while it is 0.61 for the single-task RF using only important descriptors, representing an improvement of about 24.6%. Similarly, for $\alpha$, the $R^2$ is 0.78 for the wD-MPNN compared to 0.60 for the single-task RF with important descriptors, resulting in an improvement of approximately 30%. In K, the improvement is smaller, with $R^2$ values of 0.70 for the wD-MPNN compared to 0.691 for RF models, showing an improvement of about 1.3%. For properties like $R_g$, the RF model has an $R^2$ of 0.65, compared to 0.63 for the wD-MPNN, demonstrating a minor improvement of about 3.1%.

The comparative performance of RF and wD-MPNN models highlights fundamental differences in how molecular representations capture structure-property relationships. RF models excel in predicting properties like density, $C_p$, and $C_v$ because these properties strongly correlate with well-defined molecular descriptors, which effectively capture local atomic and electronic environments. In contrast, we hypothesize that wD-MPNNs outperform RF models for properties such as $\alpha$, $\gamma$, and K due to their ability to learn hierarchical and non-local structural interactions. These properties are inherently governed by long-range connectivity, cooperative intermolecular interactions, and mechanical response behaviors that are difficult to encode in a fixed set of molecular descriptors. For instance, thermal expansion coefficients ($\alpha$, $\gamma$) and bulk modulus (K) depend not only on atomic composition but also on the spatial arrangement and flexibility of the polymer network, which wD-MPNNs can learn from graph representations of molecular topology. Unlike RF models, which rely on predefined descriptors, wD-MPNNs dynamically extract features that reflect polymer conformation, cross-linking effects, and electronic delocalization, key factors influencing macroscopic mechanical and thermal properties. This distinction explains why descriptor-based RF models are well-suited for capturing intrinsic molecular attributes, particularly in the low-data setting we are operating in for this work, whereas graph-based MPNNs excel at modeling emergent, structure-dependent behaviors. Leveraging both approaches can provide a more comprehensive framework for copolymer design, balancing accuracy and interpretability to optimize materials for targeted applications.

## Conclusion

This study evaluated the predictive capabilities of random forest (RF) and weighted, directed message passing neural network (wD-MPNN) models for seven distinct physical properties of copolymers. Our models were trained on data generated via molecular dynamics (MD) simulations, which were validated against experimental data from references, yielding $R^2$ values of 0.85 for density, 0.53 for specific heat capacity at constant pressure, 0.965 for linear expansion coefficient, 0.691 for bulk modulus, and 0.751 for volume expansion coefficient, and Spearman's rank correlation coefficients ($\rho_s$) greater than 0.8 for most properties. The results demonstrate that RF models, relying on molecular descriptors, are particularly effective for predicting density and heat capacities, where well-defined molecular features dominate. In contrast, wD-MPNN models, which leverage graph-based molecular representations, excel in predicting expansion coefficients and bulk modulus, where non-local structural interactions and polymer connectivity play a crucial role.

These insights highlight the importance of domain knowledge in choosing the appropriate molecular representation based on the underlying structure-property relationships one expects when building machine learning models in low-data settings. We found that descriptor- based RF models easily capture intrinsic molecular attributes, making them well-suited for properties explainable by local atomic environments. Conversely, graph-based wD-MPNNs can dynamically learn hierarchical interactions, making them advantageous for properties influenced by long-range connectivity and mechanical responses. Integrating both approaches offers a powerful framework for copolymer design, balancing interpretability and predictive accuracy to guide the discovery of materials with tailored properties.

# Acknowledgement

EK and MH acknowledge funding provided by the Ministerio de Ciencia, Innovación y Universidades and Agencia Estatal de Investigación in the form of FPI fellowship and the PID2023-150813OB-I00 project funded by MICIU/AEI/10.13039/501100011033 and EU's FEDER funds. RM acknowledges funding from the Wallenberg AI, Autonomous Systems, and Software Program (WASP), supported by the Knut and Alice Wallenberg Foundation.

# Data Availability Statement

The data used in this study is available free of charge on Zenodo.

Specifically, the following files are provided:

- Filename: Dataset.xlsx This file contains the information: polymer names, SMILES notation of the monomers, fraction of monomers in copolymers, and simulated and experimental values of various properties.

- Filename: Descriptors.xlsx This file provides the calculated descriptors of the copoly- mers using the PaDEL-Descriptor software.

Additionally, all code for the models and analysis developed in this work is available on GitHub.

# Supporting Information

The supporting information contains the mean squared error (MSE) performance for the training, validation, and test sets across all models. Additionally, it includes plots that compare the expected versus predicted values from the machine learning models.